\documentclass[fleqn,usenatbib]{mnras}

\usepackage{newtxtext,newtxmath}

\usepackage[T1]{fontenc}
\usepackage{ae,aecompl}
\usepackage{pdfpages}

\usepackage{graphicx}	
\usepackage{amsmath}	
\usepackage{amssymb}	

\title[FRBs from negative latency BNS with the MWA]{Using negative-latency gravitational wave alerts to detect prompt radio bursts from binary neutron star mergers with the Murchison Widefield Array}

\author[C. W. James et al.]{
Clancy W. James,$^{1}$\thanks{E-mail: clancy.james@curtin.edu.au (CWJ)}
Gemma E. Anderson,$^{1}$
Linqing Wen,$^{2}$
Joel Bosveld,$^{2}$
\newauthor
Qi Chu,$^{2}$
Manoj Kovalam,$^{2}$
Teresa J. Slaven-Blair$^{2}$
and Andrew Williams$^{1}$
\\
$^{1}$International Centre for Radio Astronomy Research, Curtin University, GPO Box U1987, Perth, WA 6845, Australia\\
$^{2}$OzGrav-UWA,  Department of Physics, The University of Western Australia, Crawley, WA 6009, Australia
}

\date{Accepted XXX. Received YYY; in original form ZZZ}

\pubyear{2019}

\begin{document}
\label{firstpage}
\pagerange{\pageref{firstpage}--\pageref{lastpage}}
\maketitle

\begin{abstract}
We examine how fast radio burst (FRB)-like signals predicted to be generated during the merger of a binary neutron star (BNS) may be detected in low-frequency radio observations triggered by the aLIGO/Virgo gravitational wave detectors. The rapidity, directional accuracy, and sensitivity of follow-up observations with the Murchison Widefield Array (MWA) are considered. We show that with current methodology, the rapidity criteria fails for triggered MWA observations above 136\,MHz for BNS mergers within the aLIGO/Virgo horizon, for which little dispersive delay is expected. A calculation of the expected reduction in response time by triggering on `negative latency' alerts from aLIGO/Virgo observations of gravitational waves generated by the BNS inspiral is presented. This allows for observations up to 300\,MHz where the radio signal is expected to be stronger. To compensate for the poor positional accuracy expected from these alerts, we propose a new MWA observational mode that is capable of viewing one quarter of the sky. We show the sensitivity of this mode is sufficient to detect an FRB-like burst from an event similar to GW170817 if it occurred during the ongoing aLIGO/Virgo third science run (O3).

\end{abstract}

\begin{keywords}
radio continuum: transients -- gravitational waves -- stars: neutron -- methods: observational
\end{keywords}

\section{INTRODUCTION}

   The detection of the first binary neutron star (BNS) merger, GW170817 \citep{2017PhRvL.119p1101A} --- also detected as GRB\,170817A \citep{2017ApJ...848L..13A,2017ApJ...848L..14G} --- triggered a wide range of follow-up observations across the electromagnetic and particle spectrum \citep{2017ApJ...848L..12A,2017ApJ...850L..35A}. However, the delay in issuing the alert prevented most instruments from observing any prompt transient event.
   
   The utility of a rapid trigger--response system for capturing electromagnetic signatures from gravitational wave (GW) events produced by compact mergers that include at least one neutron star (NS) has been long-recognised, with predictions for associated prompt radio, optical, x-ray, and gamma-ray emission \citep{1986ApJ...308L..43P,2001MNRAS.322..695H,2012ApJ...748..136C,2012IAUS..285..191C,2016MNRAS.459..121C,2019arXiv190502509R}. Consequently, the third science run, O3, of the aLIGO/Virgo GW detector network is issuing alerts for high-significance `Superevents' within minutes of detection. At least one high-significance binary NS (BNS) merger candidate has been detected.\footnote{As of 2019 July 24 (S190425z). S190510g and S190426c are also possible mergers involving a NS; S190518bb and S190524q have subsequently been retracted. See https://gracedb.ligo.org/latest/}
   
   The detection of fast radio bursts \citep[FRBs; ][]{2007Sci...318..777L,2013Sci...341...53T}, which are extragalactic radio transients of millisecond duration with an unexplained origin, also motivates the search for prompt radio emission associated with BNS mergers. While BNS cannot explain repeating FRBs, which might produce bursts at a rate $>10^4$\,Gpc$^{-3}$\, yr$^{-1}$ \citep{RaviNatureAstron2019}, it is quite possible that FRBs belong to more than one class \citep{2018NatAs...2..839C}. Indeed, the archetypal repeating FRB, 121102 \citep{2014ApJ...790..101S}, must be uncharacteristic of the population(s) as a whole \citep{2019MNRAS.486.5934J}. While the FRB rate of 200--1000\,sky$^{-1}$\,day$^{-1}$ \citep{2017AJ....154..117L} is much higher than that of BNS mergers, FRBs belong to a cosmological population \citep{2018Natur.562..386S} extending well beyond the current BNS merger detection horizon of aLIGO/Virgo. The BNS merger event rate of $1540^{+3200}_{-1220}$\,Gpc$^{-3}$\, yr$^{-1}$ \citep{2017PhRvL.119p1101A} is compatible with the non-repeating FRB rate of $\sim2700$\,Gpc$^{-3}$\, yr$^{-1}$ \citep{2019arXiv190300014L}. We use the term `FRB-like' to cover the general case where BNS are predicted to produce short-duration dispersed bursts which may or may not constitute a significant fraction of the observed FRB population.
   
   A BNS merger could produce FRB-like emission from magnetic field interactions just prior to the merger during the inspiral; the disruption of fields at the point of merger; or post-merger due to either the interaction of a relativistic jet with the interstellar medium (ISM), pulsar-like emission from a supra-massive, rapidly rotating, highly magnetised NS remnant (often referred to as a magnetar), or (if unstable) the collapse of this magnetar into a black hole \citep[e.g.][]{2000A&A...364..655U,2001MNRAS.322..695H,2013ApJ...768...63L,2013PASJ...65L..12T,2014ApJ...780L..21Z,2016ApJ...822L...7W,2019arXiv190502509R}.
   
   Low frequency ($\lesssim$300\,MHz) radio telescopes are our best hope for detecting FRB-like signals from BNS. They have large fields of view (FOV) which allow them to efficiently search for poorly localised transients \citep[e.g.][]{obenberger14,2015PASA...32...46H,2016ApJ...826L..13A,2016PASA...33...50K,anderson18,callister19pp}. Furthermore, the dispersive delay due to a distant radio signal's propagation through ionised gas in the ISM and intergalactic medium (IGM) can be minutes at low frequencies, providing extra time to re-point at a newly detected event \citep[e.g.][]{kaplan15,yancey15}. However, BNS mergers detected during the O3 run will originate from the nearby Universe, where the dispersive delays may not be large enough to compensate for delays in a trigger--response system.
   
   We therefore propose a specific observational mode of the Murchison Widefield Array (MWA) to probe for prompt FRB-like radio bursts emitted by BNS mergers, which requires using `negative-latency triggering' (triggers from detections of GW generated by the BNS inspiral) from the aLIGO/Virgo detector network during its O3 run.

\section{EXPECTED RADIO BURST DELAY}

   A key characteristic of FRBs is their dispersion sweep, being the frequency-dependent delay due to ionised gas in interstellar and intergalactic media.
   The delay, $t_{\rm FRB}$, is given by:
   \begin{eqnarray}
   t_{\rm FRB} & = & 415~{\rm DM}\,\left(\frac{\nu}{100\,{\rm MHz}}\right)^{-2}~[{\rm ms}], \label{eq:dmdelay}
   \end{eqnarray}
   where DM is the dispersion measure (total line-of-sight electron content, pc\,cm$^{-3}$), and $\nu$ the observing frequency.
   
   The BNS merger detection horizon during the O3 aLIGO/Virgo run is estimated to be 170\,Mpc, within which the event rate will be $32_{-25}^{+66}$\,yr$^{-1}$ \citep[based on BNS merger rate estimates by ][]{2017PhRvL.119p1101A}. At these distances, the contribution of the IGM to the dispersion measure \citep[$\sim$0.21\,pc\,cm$^{-3}$ Mpc$^{-1}$;][]{2004MNRAS.348..999I} will be much smaller than that due to the ISM of the Milky Way \citep[$\sim 40$\,pc\,cm$^{-3}$ at high Galactic latitudes;][]{2002astro.ph..7156C}, or that expected from its halo \citep[50--80\,pc\,cm$^{-3}$;][]{2019MNRAS.485..648P}.
   
   The observation of FRBs with DMs of 110\,pc\,cm$^{-3}$ \citep[FRB~180729.J1316+55; ][]{2019Natur.566..230C} and 114\,pc\,cm$^{-3}$ \citep[FRB~171020; ][]{2018Natur.562..386S} favour the lower limit of predictions for the halo contribution, and rule out large DM contributions from merger ejecta or the host galaxy for the majority of bursts.
    Neither of these presumably nearby FRBs occurred during one of the aLIGO/Virgo science runs.
   We therefore adopt the following DM model, applicable to the majority of FRBs originating within the current aLIGO/Virgo horizon at distance $D$:
   \begin{eqnarray}
   {\rm DM} & = & 90+0.21\,\frac{D}{\rm Mpc}~{\rm pc\,cm}^{-3}. \label{eq:dm}
   \end{eqnarray}

\section{PROPOSED OBSERVATION METHOD}
   \label{sec:obs_method}
   
   The MWA is a low-frequency (80--300\,MHz) radio telescope located in outback Western Australia \citep{2013PASA...30....7T,2018PASA...35...33W}. It has a rapid-response capability that enables it to be on-target and observing within 6--14\,s of receiving an external trigger (Hancock et al. in prep.). It is capable of triggering observations with the Voltage Capture System \citep[VCS; ][]{2015PASA...32....5T}, which has a time resolution of 100\,$\mu$s. The VCS allows for much greater sensitivity to dispersed pulses \citep[e.g.][]{2018ApJ...869..134M,2019arXiv190500598X} than can be obtained by performing image plane de-dispersion searches for prompt emission on second-timescales using data output by the standard MWA correlator \citep[e.g.][]{tingay15,2018ApJ...867L..12S}.
   
   The MWA is composed of 256 tiles --- of which 128 can be used simultaneously --- of 16 dipoles each. Beamforming on each tile is performed prior to digitisation, i.e.\ unlike LOFAR \citep{2013A&A...556A...2V}, \emph{a posteriori} beamforming with digitised data from individual dipoles is not possible. The MWA FOV is thus limited to that of an individual tile (i.e.\ its primary beam), being 610\,deg$^2$ at the peak sensitivity of 150\,MHz \citep{2017PASA...34...62S}.
   
   The standard MWA response to GW triggers uses an algorithm to maximize the overlap between tile pointing directions and GW event localization files \citep{2016PASA...33...50K}. However, it takes several seconds to download and analyse the skymaps, and the typical GW localization error regions are usually bigger than the MWA beamsize. Furthermore, localisations for negative-latency triggers will be even less accurate, being always generated at threshold (Section~\ref{sec:negative_latency}).
   
   The chance of viewing an FRB can be maximized by disabling 15 of 16 MWA dipoles on each tile, recovering the full FOV of a single dipole (the entire visible sky). Since dipole sensitivity tapers towards the horizon, we characterise the FOV in this mode as $\pi$\,sr, i.e.\ above an elevation of $30$\,deg, viewing $1/4$ of all BNS mergers. Compared with MWA VCS observations using all 16 dipoles per tile, the loss of sensitivity will be approximately 16-fold. However, BNS mergers detected by aLIGO/VIRGO will be significantly closer than the majority of observed FRBs, largely compensating for this loss of sensitivity.
   
    We propose to trigger MWA's VCS whenever an aLIGO/Virgo template search with at least one mass consistent with a NS exceeds a pre-set threshold. Observations lasting a single minute with the VCS using a single dipole per tile will be adequate to catch the majority of proposed FRB-like signals predicted to be produced during a merger. Highly-dispersed FRBs from the centres of local galaxies/clusters, or those propagating through much of the Milky Way's disc, may have DMs over $\sim$1000\,pc\,cm$^{-3}$ \citep{2019MNRAS.485..648P}. These will be observable using the methods of \citet{2016PASA...33...50K}, which will also be sensitive to bursts emitted by any post-merger remnant.

   \section{NEGATIVE-LATENCY TRIGGERING}
   \label{sec:negative_latency}
   
   The time between the aLIGO/Virgo detection of a Superevent during the O3 run and the submission time of alerts is currently 18--28\,s.\footnote{https://gracedb.ligo.org/latest/} Let us suppose we use MWA to trigger on the first BNS merger detected by any GW pipeline, rather than waiting for the most significant trigger over all pipelines, and that this alert is broadcast at the best alert time of 18s. Adding the typical MWA trigger response time of $t_{\rm MWA}=10$\,s, we estimate a total delay between merger and observation of $t_{\rm obs}=28$\,s.
   
   Using equations~(\ref{eq:dmdelay}) and (\ref{eq:dm}), an observational delay of $t_{\rm obs}=28$\,s matches the dispersive delay of a 136\,MHz signal generated at at 170\,Mpc (DM=125.7\,pc\,cm$^{-3}$), i.e.\ the maximum possible observing frequency that will allow us to observe an FRB-like signal associated with a BNS merger is $\nu_{\rm max}=136$\,MHz. At the 40\,Mpc distance of GW170817 \citep{PhysRevLett.119.161101,2019PhRvX...9a1001A}, $\nu_{\rm max}$ is even lower, at 121\,MHz (DM=98.4\,pc\,cm$^{-3}$).
   
   Despite evidence from GHz observations that FRB emission is stronger with decreasing frequency \citep{2019ApJ...872L..19M}, MWA observations at 170--200\,MHz did not detect seven FRBs discovered by the Australian Square Kilometre Array Pathfinder \citep[ASKAP; ][]{2008ExA....22..151J} during simultaneous observations \citep{2018ApJ...867L..12S}. As discussed by \citet{2018ApJ...867L..12S}, this suggests a low-frequency downturn in the spectral strength, possibly due to free-free absorption \citep{1986rpa..book.....R}, e.g.\ in merger ejecta. Scattering due to inhomogeneities in the ISM \citep{2004ApJ...605..759B} and/or the IGM \citep{2013ApJ...776..125M} would also spread the signal in time and reduce sensitivity, although this alone cannot explain the non-detection. Given FRBs have been observed at 400\,MHz by \citet{2019Natur.566..230C}, a complete lack of $\sim$100\,MHz emission seems unlikely. Furthermore, bursts emanating from within the aLIGO/Virgo horizon will be at least an order of magnitude closer, and therefore brighter, than those observed by ASKAP \citep{2018Natur.562..386S}. In order to catch these events at higher MWA frequencies (136--300\,MHz), where any signal is likely to be stronger, we require negative-latency triggering.
   
   \subsection{Simulations using GW170817}
   
   The principle of negative-latency triggering is simple: search for GWs produced by the in-spiral of compact objects prior to their merger, and broadcast the alert as soon as the significance of a template waveform passes a pre-defined threshold, rather than waiting for the merger and the maximum sensitivity of a template to be reached \citep{2012ApJ...748..136C,PhysRevD.85.102002}. It is specifically planned to be implemented in GW search algorithms, such as the Summed Parallel Infinite Impulse Response {\sc SPIIR} pipeline \citep{PhysRevD.86.024012,001e1fde32aa46b18aee58f4216dfd13,PhysRevD.85.102002,7498910c2c4f45b59d657d626ea2a1f9,76e97cf9801544919973534ed7028b6a,2018CoPhC.231...62G}.
   
   We analyse the trade-off in sensitivity and time using publicly available 2048\,Hz clean GW data on GW170817 from the LIGO Hanford (`H') and Livingston (`L') detectors (the Virgo signal-to-noise ratio, SNR, is negligible here compared to the SNRs of the two LIGO detectors) \citep{2015JPhCS.610a2021V}.\footnote{https://www.gw-openscience.org/events/GW170817/} A time-domain template waveform is generated in {\sc PyCBC} \citep{2005PhRvD..71f2001A,2012PhRvD..85l2006A,2014PhRvD..90h2004D,2017ApJ...849..118N,alex_nitz_2019_2643618} using the {\sc SpinTaylorT4} approximant, with a parameter set within the range of the best-fit parameters found by \citet{2017PhRvL.119p1101A,2018PhRvL.121p1101A}. The background power spectral density is estimated from the data prior to merger. A low-frequency cut-off of 20\,Hz was applied.
   
\begin{figure}
   \centering
   \includegraphics[width=\columnwidth]{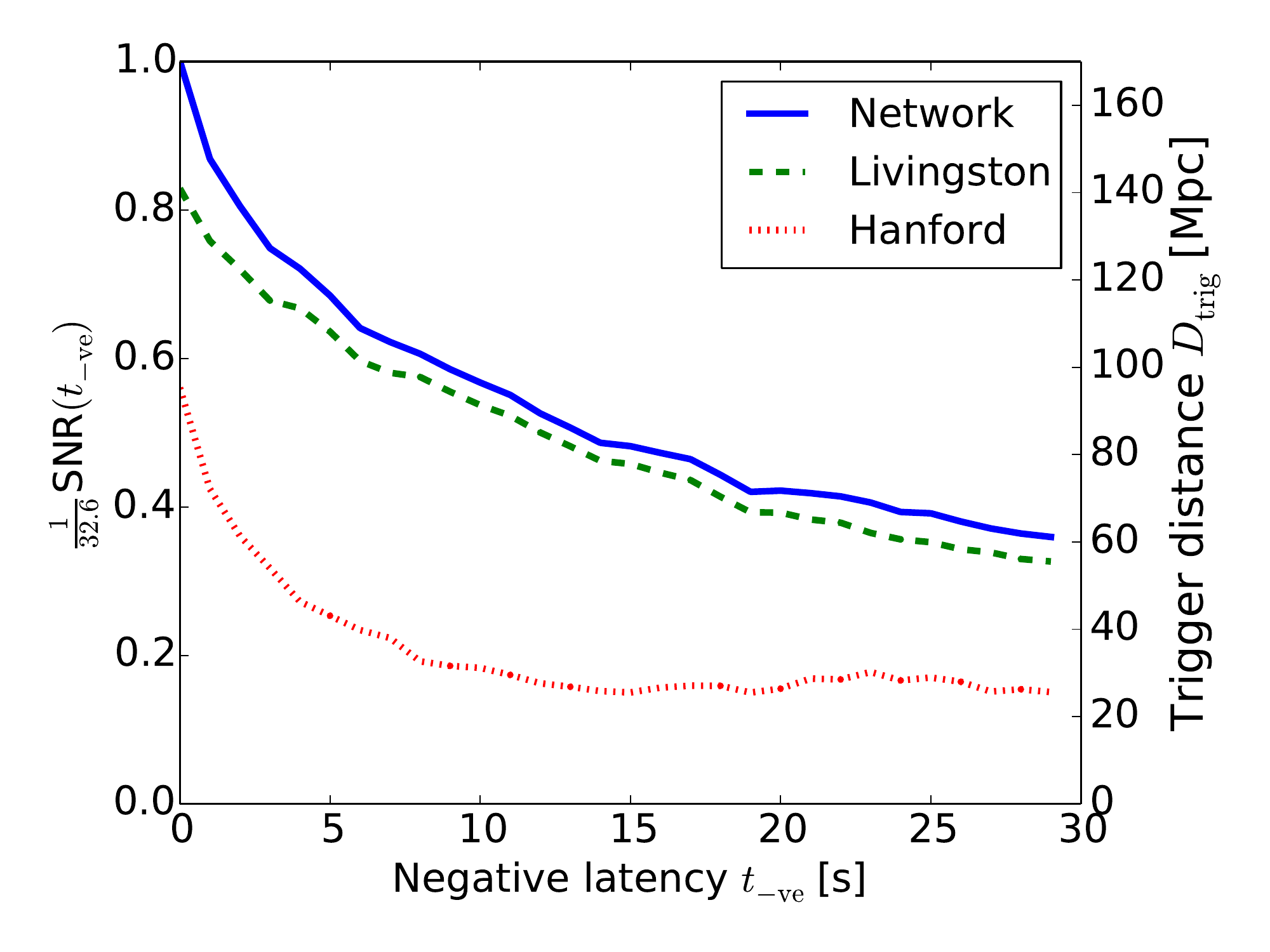}
      \caption{Signal-to-noise ratios (SNR; left axis) of a negative-latency trigger applied to GW170817 as a function of time ($t_{\rm -ve}$) prior to the merger. Shown are the SNRs from the Livingston and Hanford detectors, and the network SNR from equation~(\ref{eq:coherent_snr}). It has been normalised by the peak network SNR value of $32.6$ (at $t_{\rm -ve}=0$) to highlight the loss of sensitivity with increasing negative latency. The network SNR also gives the corresponding distance ($D_{\rm trig}$; right axis) at which a GW170817-like event would equal the detection threshold at a time $t_{\rm -ve}$ in the aLIGO/Virgo O3 run.
              }
         \label{fig:snr_dist}
   \end{figure}

   Negative-latency triggering is simulated by setting the predicted waveform shape to zero from a time $t_{\rm -ve}$ prior to the merger onwards. The network signal-to-noise at $t_{\rm -ve}$, SNR($t_{\rm -ve}$), is calculated from the individual SNRs on each detector as:
   \begin{eqnarray}
   {\rm SNR}(t_{\rm -ve}) =  \sqrt{{\rm SNR}_{\rm L}^2(t_{\rm -ve})+{\rm SNR}_{\rm H}^2(t_{\rm -ve})}. \label{eq:coherent_snr}
   \end{eqnarray}
   This is shown in Fig.~\ref{fig:snr_dist}, normalised by the peak value of 32.6 found for the SNR of GW170817 at $t_{\rm -ve}=0$. This peak SNR is in close agreement with that found by aLIGO/VIRGO \citep{2017PhRvL.119p1101A}. Over the range $t_{\rm -ve}\in 0$--$30$\,s, the SNR drops to one third of its peak value.

   Nearby BNS mergers will produce a stronger GW signal, resulting in the SNR passing the trigger threshold at earlier times. Conversely, events that produce a negative-latency trigger at an earlier $t_{\rm -ve}$ will have to be closer. We therefore define the trigger distance, $D_{\rm trig}(t_{\rm -ve})$, to be the distance at which a GW170817-like event would cause a trigger at time $t_{\rm -ve}$. Since the SNR of a GW event scales inversely with distance, $D_{\rm trig}(t_{\rm -ve})$ will be directly proportional to SNR$(t_{\rm -ve})$. The constant of proportionality is set by the maximum BNS detection distance for the O3 run of $D_{\rm max}=170$\,Mpc at $t_{\rm -ve}=0$. $D_{\rm trig}$ is shown in Fig.~\ref{fig:snr_dist} using the same curves as SNR$(t_{\rm -ve})$ via the right-hand y-axis.

\begin{figure}
   \centering
   \includegraphics[width=\columnwidth]{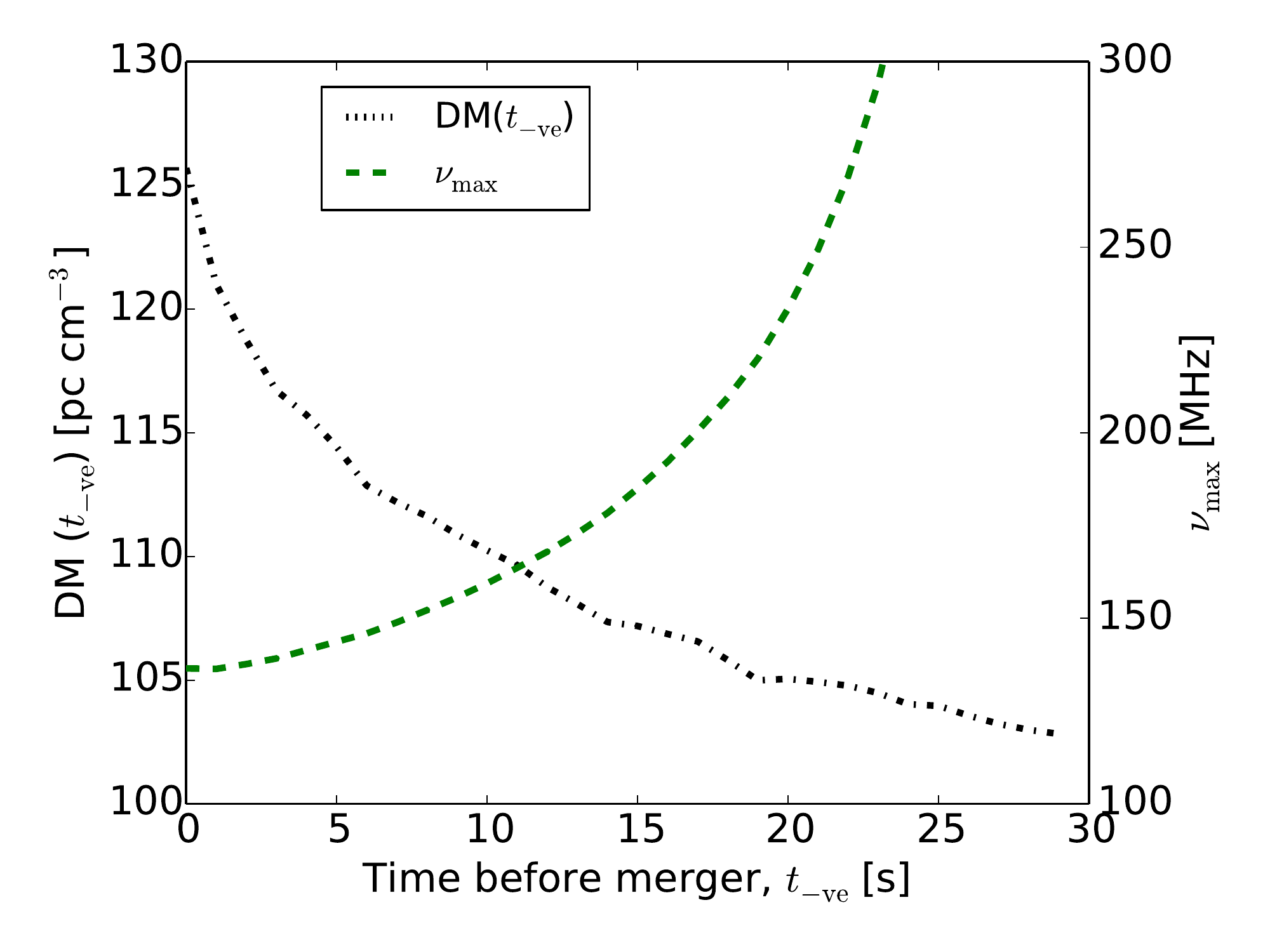}
      \caption{Expected dispersion measure, DM, for FRB-like events associated with BNS mergers as a function of $t_{\rm -ve}$ (left axis, black dotted line). For an assumed total observational delay $t_{\rm obs}=28$\,s, the corresponding peak frequency ($\nu_{\rm max}$) at which an FRB-like signal would be observable within the GW-MWA response time of $28$\,s (right axis, green dot-dashed line).
              }
         \label{fig:dm_freq}
\end{figure}

The distance of an event is associated with a dispersion measure (DM) via equation~(\ref{eq:dm}). Applying this to $D_{\rm trig}$, itself a function of $t_{\rm -ve}$, produces DM$(t_{\rm -ve})$, shown in Fig.~\ref{fig:dm_freq}. That is, a BNS merger passing the detection threshold at earlier times must be closer, and hence will have passed through less intervening material (and vice versa).

 For a given frequency $\nu$, the time delay $t_{\rm FRB}({\rm DM},\nu)$ in the arrival of an FRB-like signal due to dispersion can be found through equation~(\ref{eq:dmdelay}). In order to observe the event, the total response time ($t_{\rm obs}$), less the time gained in negative latency ($t_{\rm -ve}$), must be less than the dispersed arrival time of the FRB ($t_{\rm DM}(\nu)$). Equivalently, there is a maximum observable frequency ($\nu_{\rm max}$) for which the above statement holds true, i.e.:
\begin{eqnarray}
t_{\rm FRB}({\rm DM},\nu_{\rm max}) & = & t_{\rm obs} - t_{\rm -ve}, \label{eq:numax}
\end{eqnarray}
Solving equation~(\ref{eq:numax}) using the values of DM$(t_{\rm -ve})$ given in Fig.~\ref{fig:dm_freq} sets $\nu_{\rm max}$ to the maximum possible observing frequency for which the MWA can be on target to observe an FRB-like signal associated with a merger as a function of negative latency. While nearby BNS mergers incur less dispersive delay, this is more than compensated-for by the time gained through negative-latency triggering. Thus $\nu_{\rm max}$ increases from $136$\,MHz at $t_{\rm -ve}=0$ to infinity when $t_{\rm -ve}=t_{\rm obs}$, where the dispersion delay ($t_{\rm FRB}({\rm DM},\nu_{\rm max})$) is no longer required to enable a follow-up observation. For the MWA, this maximum frequency is capped at 300\,MHz.

Fig.~\ref{fig:dm_freq} suggests that triggered observations should use a tuneable frequency band based on the negative latency of the trigger. As the exact relation between $t_{\rm -ve}$ and $\nu_{\rm max}$ will depend on the shape of the GW signal, we suggest using a split frequency band (e.g.\ 112--127\,MHz, and 216--231\,MHz). This provides a good compromise between greater detection sensitivity at high frequencies to nearby events, and a greater dispersive delay at low frequencies allowing more-numerous distant events to be observed.

Given the lack of definitive predictions for the radio strength of FRB-like signals from BNS mergers, it is currently impossible to translate the gains from negative-latency triggering into a hard prediction for the detection rate of FRB-like signals from BNS mergers. Rather, our prescription is to use a negative-latency trigger, allowing observations at the highest possible frequency, to maximize the chance of observing any such burst.

We can demonstrate that our MWA observational mode proposed in Section~\ref{sec:obs_method} is sufficiently sensitive to detect FRB-like emission from BNS mergers in at least one plausible scenario. Assuming that BNS mergers are responsible for a significant fraction of the detected non-repeating FRB population, it becomes possible to use the (poor) constraints on this population to make event-rate predictions. For example, in supplemental material, we present a calculation based on the properties of the observed FRB population \citep{2018Natur.562..386S,2019ApJ...872L..19M,2019arXiv190300014L}. For the distance estimated by \citet{2018ApJ...867L..10M}, FRB~171020 places the strongest constraints on FRB emission in the MWA band \citet{2018ApJ...867L..12S}. Using this to set an absolute scale for emission strength at MWA frequencies, our conclusion \emph{for this particular calculation} is that all FRB-like bursts produced by BNS mergers within 30\,Mpc will be detectable with the MWA if they are located within the $\pi$\,sr observable sky with this mode. At larger distances, there is a decreasing probability that FRB-like signals will be sufficiently bright to be detectable. For example, GW170817 occurred at a distance of 40\,Mpc, corresponding to an estimated 80\% probability of producing a detectable FRB-like signal. If a similar event were to occur during the aLGIO/Virgo O3 run, our proposed observational mode would either detect an FRB-like burst, or place strong constraints on BNS mergers as FRB progenitors.

\section{CONCLUSION}

The detection of prompt FRB-like signals associated with BNS mergers requires a radio instrument of sufficient sensitivity capable of being on-target before the arrival of a burst. We have described a unique observational strategy whereby candidate BNS GW events are identified with negative latency (i.e.\ prior to merger) by aLIGO/Virgo, and trigger automatic and rapid MWA follow-up observations. This gain in response time would allow us to trigger higher-frequency observations (up to 300\,MHz) with the MWA, where any burst signal is less likely to suffer from scatter broadening or free-free absorption.

Other low-frequency radio telescopes, such as the Low Frequency Array \citep[LOFAR; ][]{2013A&A...556A...2V} and the Long Wavelength Array \citep[LWA; ][]{2009IEEEP..97.1421E}, may also be able to take advantage of negative-latency triggers broadcast by aLIGO/Virgo, improving global sky-coverage.

For a plausible model of FRB-like emission from BNS mergers (see supplemental material), we have shown that this observational method is sufficiently sensitive to allow for a detection. This proposed experiment presents the best, and perhaps only, chance of testing whether FRB-like signals are produced during a BNS merger, \emph{and is feasible during the O3 run of aLIGO/Virgo}.

\section*{Acknowledgements}

GEA is the recipient of an Australian Research
Council Discovery Early Career Researcher Award (project number DE180100346). This work was supported in part by the Australian Research Council Centre of Excellence for Gravitational Wave Discovery
(OzGrav; project number CE170100004).
This research has made use of data, software and/or web tools obtained from the Gravitational Wave Open Science Center (https://www.gw-openscience.org), a service of LIGO Laboratory, the LIGO Scientific Collaboration and the Virgo Collaboration. LIGO is funded by the U.S. National Science Foundation. Virgo is funded by the French Centre National de Recherche Scientifique (CNRS), the Italian Istituto Nazionale della Fisica Nucleare (INFN) and the Dutch Nikhef, with contributions by Polish and Hungarian institutes. Calculations in this work use NumPy \citep{NumPy} and Matplotlib libraries \citep{Hunter:2007}.

\bibliographystyle{mnras}
\bibliography{bibliography}

\bsp	
\label{lastpage}



\section*{Supplementary material}

# S1 SUPPLEMENTAL MATERIAL — SENSITIVITY ESTIMATE

We present an estimate of the sensitivity of our proposed MWA observational mode to a plausible scenario for FRB-like signals associated with BNS mergers. Rather than beginning with one of the theoretical predictions for FRB-like emission from BNS mergers discussed in Section 1, and applying the effects of propagation through any ejecta in the merger environment, our approach is to assume that BNS mergers are responsible for the majority of the FRB population. This allows us to use the observed properties of this population to model FRB-like emission from BNS mergers, and in particular observations of FRB 171020 (Shannon et al. 2018; Sokolowski et al. 2018), since this event places the strongest limit on FRB emission at MWA frequencies. Here we show that our proposed MWA observational mode, when combined with negative-latency triggers from aLIGO/Virgo, is sensitive to such a model, and therefore sensitive enough to either discover FRB-like bursts from BNS mergers or rule out an entire class of FRB progenitor models.

To generate a model of the radio fluence of FRB-like bursts associated with BNS mergers, we first scale the properties of FRB 171020 with the frequency, distance and probability distribution of burst energies of the larger FRB population. We then estimate the sensitivity of the MWA to this model when using only one in 16 dipoles on each tile, and calculate the probability of an FRB-like burst from a BNS merger being detected as a function of distance to the event. We conclude with an estimate of the total event rate, and recommend a particular choice of observation frequencies for the MWA.

We caution that the resulting estimate is based on parameters of the FRB population that are poorly constrained, and we do not attempt to present uncertainty estimates on our results. Suffice to say they will be large. Furthermore, these estimates depend on the results derived in the main manuscript in Section 4 using GW170817. Specifically, the relation between the distance ($D_{\rm trig}$) at which a BNS merger will generate a negative-latency trigger, and the resulting maximum possible observing frequency ($\nu_{\rm max}$).

These results should therefore be interpreted qualitatively, i.e. that if BNS mergers produce associated FRB-like bursts constituting some part of the observable FRB population, our proposed observational mode will be able to detect them in some fraction of the aLIGO/Virgo O3 sensitive volume.

## S1.1 Properties of FRB 171020

Sokolowski et al. (2018) used the MWA to perform simultaneous (also known as shadowing) observations of ASKAP during the Commensal Real-Time ASKAP Fast-Transients (CRAFT) Survey at 1,296 MHz (Bannister et al. 2017; Shannon et al. 2018). Despite evidence that the FRB spectral strength increases at lower frequencies (Macquart et al. 2019), none of the seven FRBs detected by ASKAP during this period were detected by the MWA.

Using 0.5 s-duration dedispersed images, Sokolowski et al. (2018) obtained upper limits on the fluence at 184 MHz, $\mathcal{F}_{184}$, of three of the seven FRBs: 2200, 3350, and 450 Jy ms for FRBs 171020, 180110, and 180324 respectively. Of these, FRB 171020 is likely to be closest, with a measured DM of 114.1 pc cm$^{-3}$, and a possible host galaxy identified at a luminosity distance of 37 Mpc (Mahony et al. 2018). This FRB also lies in the low-energy range of the DM–fluence relation identified by Shannon et al. (2018), which implies radio powers at GHz frequencies of $10^{30}$–$10^{34}$ erg Hz$^{-1}$. We therefore use the limiting fluence of 2200 Jy ms for FRB 171020 to define a minimum 184 MHz fluence, $\mathcal{F}_{\rm min}^{184}$ = 2200 Jy ms, for bursts at 37 Mpc, corresponding to $3.6 \cdot 10^{30}$ erg Hz$^{-1}$ at 37 Mpc.

## S1.2 Frequency, distance, and burst energy distribution

The cumulative distribution $p(E > E_0)$ of FRB energies $E$ above some value $E_0$ can be described by a power-law of the form:

$$p(E > E_0) = \left(\frac{E_0}{E_{\rm min}}\right)^{\gamma}, \quad (S1)$$

where $E_{\rm min}$ is the minimum energy, and $\gamma = -0.7$ is the power-law index (Lu & Piro 2019). In the local Universe, the fluence distribution will be proportional to the energy distribution, so that the probability, $p_{\rm det}$, of detecting a burst above a frequency-dependent fluence threshold, $\mathcal{F}_{\rm th}(\nu)$, becomes:

$$p_{\rm det} = \left(\frac{\mathcal{F}_{\rm th}(\nu)}{\mathcal{F}_{\rm min}(\nu, D)}\right)^{\gamma}. \quad (S2)$$

Here, $\mathcal{F}_{\rm min}$ is the minimum fluence expected for an FRB at frequency $\nu$ and distance $D$. We model $\mathcal{F}_{\rm min}$ by scaling the fluence upper limit of FRB 171020 by distance, $\mathcal{F} \propto D^{-2}$. Using the free-free absorption model of Sokolowski et al. (2018), the optical depth at 184 MHz found for FRB 171020 is $\tau_{\rm ff}^{184} = 1.24$ (Sokolowski et al. 2018). Scaling the optical depth as $\tau_{\rm ff} \propto \nu^{-2.1}$ (characteristic of free-free absorption) produces the following model for $\mathcal{F}_{\rm min}(\nu, D)$:

$$\mathcal{F}_{\rm min}(\nu, D) = \mathcal{F}_{\rm min}^{184} e^{-\tau_{\rm ff}^{184}\left[\left(\frac{\nu}{184 \text{ MHz}}\right)^{-2.1}-1\right]} \left(\frac{D}{37 \text{ Mpc}}\right)^{-2}. \quad (S3)$$

## S1.3 Detection threshold

The expected radio signal-to-noise ratio (SNR$_{\rm FRB}$) for a transient source of fluence $\mathcal{F}$ is determined by the radiometer equation:

$$\text{SNR}_{\rm FRB}(\mathcal{F}) = \frac{\mathcal{F}}{\text{SEFD}}\sqrt{\frac{\Delta\nu}{2\Delta t}}, \quad (S4)$$

for bandwidth $\Delta\nu$ (here, 30.72 MHz) and burst duration $\Delta t$. The factor of '2' accounts for two polarisation channels. SEFD is the system equivalent flux density, which can be calculated from the effective areas and system temperatures given for individual MWA dipoles (i.e. one of every 16 on a tile) up to 220 MHz by Wayth et al. (2017). We perform a quadratic fit to extrapolate the SEFD for frequencies in the 220–300 MHz range. Since the SEFD is inversely proportional to the number of independent elements, we divide the single-dipole SEFD by the number of MWA tiles that can be used simultaneously, i.e. 128. This gives a minimum SEFD of 3,670 Jy at 168 MHz, increasing to 13,200 Jy at 350 MHz.

We scale the burst width of FRB 171020, $\Delta t$, from its

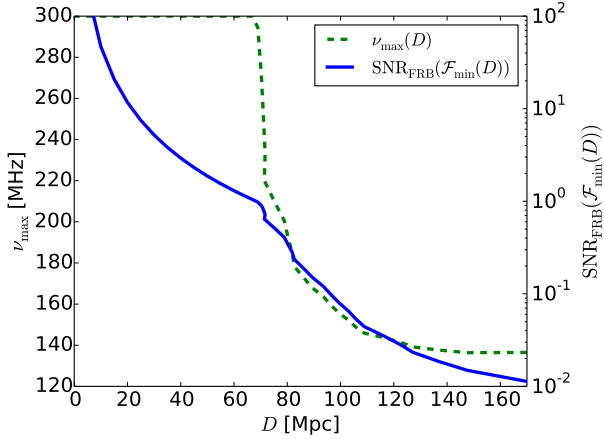

**Figure S1.** Green dashed line (left y-axis): maximum observation frequency ($\nu_{max}(D)$), calculated as a function of BNS merger distance $D$ from Figs. 1 and 2 in the main manuscript. Blue solid line (right y-axis): minimum FRB SNR ($SNR_{FRB}$) as a function of distance $D$, assuming MWA observations at $\nu_{max}$, and a minimum fluence ($\mathcal{F}_{min}$) given by equation (S3).

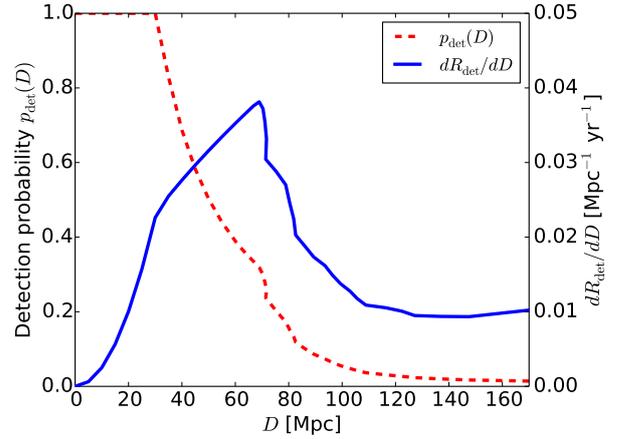

**Figure S2.** Red dashed line (left y-axis): detection probability $p_{det}$ of FRBs above a threshold of $SNR_{FRB,th} = 5$ as a function of distance $D$, accounting for the distribution of FRB fluences above $\mathcal{F}_{min}$ and observations at $\nu = \nu_{max}$ as shown in Figure S1. Blue solid line (right y-axis): differential rate $dR_{det}/dD$ of detecting FRB-like bursts from BNS mergers as a function of event distance $D$. Green dashed line (right y-axis): maximum observation frequency ($\nu_{max}(D)$), calculated from Figs. 1 and 2 in the main manuscript.

observed value at 1,184 MHz of 1.7 ms (Shannon et al. 2018) as $\Delta t \sim \nu^{-4}$ (Bhat et al. 2004), which is characteristic of scattering in the ISM (Bhat et al. 2004). The required detection threshold ($SNR_{FRB,th}$) for a statistically significant detection of a dispersed burst in MWA data depends on the range of potential positions that need to be searched. If the location of the BNS merger is well-constrained from GW or $\gamma$-ray observations, as with GW 170817, $SNR_{FRB,th} = 5$ is typical, as was used by Sokolowski et al. (2018).

### S1.4 Probability of detection

For a constant instrumental sensitivity, the model presented in equation (S3) predicts that the SNR of an FRB, $SNR_{FRB}$, at a given distance $D$ will always be maximised by observing at the highest possible frequency allowed by the dispersive delay, $\nu_{max}(D)$, defined in Section 4, equation (4). This is also the case when considering the declining sensitivity of MWA dipoles above 170 MHz (Wayth et al. 2017). The dependence of $\nu_{max}$ on $D$ is shown in Fig. S1.

Inserting $\nu = \nu_{max}(D)$ into equations (S2)–(S4) gives $SNR_{FRB}$ for the minimum fluence $\mathcal{F}_{min}$ as a function of distance $D$ in Fig. S1. The inflection point near 70 Mpc is due to the top of the MWA band (300 MHz) being reached.

Within 30 Mpc, $SNR_{FRB}(\mathcal{F}_{min})$ exceeds the detection threshold of $SNR_{FRB,th} = 5$, and all FRB-like bursts will be detectable. At greater distances, the probability of detection will be given by equation (S2), which accounts for most FRBs being intrinsically brighter than FRB 171020. This is shown in Fig. S2. The detection probability rapidly drops from 100% at 30 Mpc to below 10% for distances above 80 Mpc, although some fraction of very bright FRB-like bursts from BNS mergers will still be detectable out to 170 Mpc.

### S1.5 Absolute detection rate and choice of frequency

The total detection rate of FRB-like bursts from BNS mergers, $R_{det}$, can be calculated from $p_{det}$ via:

$$R_{det} = \int_0^{170\,\text{Mpc}} dD\, 4\pi D^2 p_{det} \Phi_0 \qquad (S5)$$

where $\Phi_0$ is the volumetric rate of BNS mergers. Given BNS merger event rate estimates of $1540^{+3200}_{-1220}\,\text{Gpc}^{-3}\,\text{yr}^{-1}$ (Abbott et al. 2017) and FRB rate estimates of $2700\,\text{Gpc}^{-3}\,\text{yr}^{-1}$ (Lu & Piro 2019), we take $\Phi_0 = 2000\,\text{Gpc}^{-3}\,\text{yr}^{-1}$. For the estimates of $p_{det}$ in Fig. S2, we estimate $R_{det} = 2.8\,\text{yr}^{-1}$ for full-sky coverage, or $0.7\,\text{yr}^{-1}$ for our MWA observational mode with a FOV of $\pi$ sr.

The differential detection rate, $dR_{det}/dD$, is given by the integrand of equation (S5). This is plotted in Fig. S2. It rises quadratically with $D$ until 30 Mpc, beyond which only a fraction of FRB-like bursts are detectable. At 70 Mpc, the GW signal of a BNS merger becomes sufficiently weak that the time gained through negative-latency triggering, $t_{-ve}$, is too small to allow observations at the maximum MWA frequency of 300 MHz. To compensate, the maximum observation frequency, $\nu_{max}$ (from Fig. S1), must be reduced, so that the increased dispersive delay offsets the response time of the trigger–alert system. Since sensitivity decreases with frequency, this causes a rapid drop in the differential detection rate. Beyond 100 Mpc, $\nu_{max}$ decreases slowly, and the effect of increasing volume balances the drop in detection probability ($p_{det}$).

Comparing the maximum frequency $\nu_{max}$ with the differential detection rate $dR_{det}/dD$ in Fig. S2 suggests an optimal choice of frequencies. Observations with the MWA allow an arbitrary choice of frequency bands totaling 30.72 MHz bandwidth within the observable range, and we propose to

observe with two bands of 15.36 MHz each. A low-frequency band from 112–127.36 MHz (just above FM frequencies) targets bursts from distances beyond 70 Mpc, while a high-frequency band from 215.68–231.04 MHz maximises the sensitivity to FRB-like bursts originating within 70 Mpc. Both bands fall within the standard frequency range used in most MWA observations (e.g. Wayth et al. 2015; Hurley-Walker et al. 2017).

This paper has been typeset from a T<sub>E</sub>X/LAT<sub>E</sub>X file prepared by the author.



\end{document}